\documentclass[aps,prd,preprint,groupedaddress,showpacs]{revtex4-1}

\usepackage{amssymb,amsmath}
\usepackage[dvips]{graphicx}

\begin{document}

\title{Renormalization of the 2PI Hartree-Fock approximation on de Sitter background in the broken phase}

\author{Takashi Arai}
\email[]{araitks@post.kek.jp}

\affiliation{KEK Theory Center, Tsukuba, Ibaraki 305-0801, Japan}
\affiliation{Graduate School of Mathematics, Nagoya University, Nagoya 464-8602, Japan}

\begin{abstract}
The infrared effects for light minimally coupled scalar fields with quartic self-interaction in de Sitter space is investigated using the 2PI effective action formalism. This formalism partially resums infinite series of loop diagrams, and enables us to circumvent the IR divergence problem for a massless minimally coupled scalar field in de Sitter space. It is anticipated that nonperturbative infrared effects generate a curvature-induced mass and self-regulate the IR divergence. However, due to its nonperturbative nature, the renormalization prescription is a nontrivial task. To calculate physical quantities, an appropriate renormalization prescription is required. In this paper, we will show that the MS-like scheme is possible at the Hartree-Fock truncation of the 2PI effective action, and infinite series of divergent terms are needed as counterterms. The phase structure and the quantum backreaction to Einstein's field equation are calculated.
\end{abstract}


\maketitle

\section{Introduction}
The study of quantum field theory in de Sitter space has a long history. The reason for this is the maximal spacetime symmetry characterized by the de Sitter group. Thanks to this symmetry, analytical expressions for the free field propagator and the one-loop effective action can be obtained~\cite{Candelas}. Furthermore, it is cosmologically relevant in the sense that our universe is believed to have undergone an inflationary expansion phase at an early stage which can be approximately described using de Sitter space with appropriate coordinates. Recently, quantum field theory in de Sitter space has again been attracting attention. This is due to the discovery of the accelerating expansion of the present universe and the temperature fluctuations in the Cosmic Microwave Background radiation. These fluctuations are assumed to be partly generated by a quantum fluctuation during the inflationary stage.

However, a problem arises when considering quantum field theory in de Sitter space: the infrared divergence for a massless minimally coupled scalar field~\cite{Allen1,Allen2}. In the course of the construction of a realistic cosmological model, the infrared divergence is usually circumvented by introducing a low-momentum cutoff. The cutoff prescription corresponds to the consideration of a local de Sitter geometry. The local de Sitter geometry is physically realistic because the inflationary epoch ceases at a finite time. In addition, it is sufficient for an effective field theory without higher order quantum corrections. However, when we take the higher order quantum corrections into account, the infrared cutoff causes difficulties in our calculation. The low-momentum infrared cutoff partially breaks the de Sitter symmetry, and gives rise to a time dependent term in the propagator~\cite{Allen1,Allen2}. This fact means that physical quantities may gain a time dependence, which grows with time, and eventually the perturbative approximation breaks down. In contrast, there have been some attempts to utilize the breaking time dependence to provide an explanation for dark energy being the vacuum expectation value of the energy momentum tensor, or the time evolution of the cosmological constant~\cite{Woodard1,Woodard2,Kitamoto}. 

On the other hand, the purely theoretical issue of the behavior of a massless field on the full de Sitter geometry remains open. A first step in studying this is to consider the resummed effects of loop diagrams. The stochastic approach was devised for the study of these loop effects~\cite{Starobinsky} where low-momentum infrared modes were treated stochastically. It was shown that the interaction generates the effective mass~\cite{Starobinsky,Riotto}. Recently, the $O(N)$ model was investigated using the $1/N$ expansion, and again the curvature-induced mass was obtained~\cite{Serreau}. More recently, using the two-particle irreducible (2PI) resummation technique at the level of the Hartree truncation, we explicitly demonstrated the generation of mass in a full quantum mechanical treatment of $\phi^4$ theory~\cite{Arai}.

The 2PI formalism is a variant of the one-particle irreducible (1PI) effective action formalism, where two-particle irreducible vacuum diagrams are used in the expansion of the effective action instead of one-particle irreducible vacuum diagrams~\cite{Jackiw}. Nonperturbative quantum loop contributions are resummed into these 2PI diagrams. Furthermore, the 2PI formalism contains the commonly used Hartree-Fock and large-$N$ approximations~\cite{Ramsey}. In the last few years, the 2PI formalism has been applied to the study of phase transitions at finite temperature and nonequilibrium quantum field theory. There, a nonperturbative treatment is required because the contributions of the temperature dependence can overwhelm the small coupling constant. However, renormalization in this formalism is a nontrivial task due to its nonperturbative nature. In fact, it often happens that different renormalization prescriptions give different results in the nonperturbative resummation scheme~\cite{Stevenson,Paz,Rischke}. In light of the BPHZ renormalization scheme, it seems that the resummation formalism requires infinitely many divergent terms as counterterms. Recently, much progress has been made in the 2PI renormalization for scalar field theory with quartic self-interaction~\cite{Blaizot1,Blaizot2}. In particular, the 2PI renormalization for any truncation order at zero temperature is investigated in detail in Ref.~\cite{Berges}. Furthermore, an explicit counterterm construction for more complicated models at the Hartree-Fock truncation of the 2PI effective action are shown in Ref.~\cite{Fejos}. 

In our previous paper, the nonperturbative infrared effects for $\phi^4$ scalar field theory were studied, with the assumption that the $\overline{\mathrm{MS}}$ scheme is possible. In this paper, we elaborate on the renormalization prescription from our previous analysis. We show that the MS-like scheme is possible in the 2PI Hartree-Fock approximation. We derive counterterm equations for the MS-like scheme, and solving these equations, we obtain expressions for the counterterms. Furthermore, the effects of the generation of mass on the phase structure and the vacuum expectation value of the energy-momentum tensor are investigated.

This paper is organized as follows. In section II, we review the 2PI effective action formalism. In particular, we show how the independent counterterms emerge for consistent renormalization. In section III, we make a renormalization program at the Hartree truncation level of the 2PI effective action in flat space. For consistent renormalization, it turns out that infinite series of divergent terms are needed as counterterms. In section IV, we extend our renormalization prescription to de Sitter space. De Sitter space partially contains the same divergence structure as flat space, and to renormalize it the same counterterms as in the case of flat space are used. We calculate the effective potential and the vacuum expectation value of the energy-momentum tensor in section V. Section VI is devoted to the conclusion. In this paper, we adopt the unit system of $c=\hbar=1$.

\section{2PI Hartree-Fock approximation}
In this section, we review the 2PI effective action formalism for $\phi^4$ scalar field theory in flat spacetime in order to designate our notations and conventions. For a single scalar field theory, the 2PI effective action which is a functional of the vacuum expectation value of the quantum field $v$ and the full propagator $G$, is given by~\cite{Jackiw}
\begin{equation}
\Gamma[v,G]=S[v]+\frac{i}{2} \log \mathrm{det} [G^{-1}]+\frac{i}{2} \! \int \! \! d^4x  \! \int \! \! d^4 x' G_0^{-1} [v](x,x') G(x',x)+\Gamma_2[v,G],
\end{equation}
where 
\begin{equation}
i G_0^{-1}[v](x,x')= \frac{\delta^2 S[v]}{\delta \phi(x) \delta \phi(x')}, 
\end{equation}
is an inverse propagator.
$\Gamma_2[v,G]$ is expressed by $(-i)$ times all of the two-particle irreducible vacuum diagrams with a propagator given by $G$ and vertices given by a shifted action $S_{\mathrm{int}}$, defined by
\begin{equation}
S_{\mathrm{int}}[\varphi]=\sum_{n=3}^{\infty}\frac{1}{n!} \left( \prod_{i=1}^{n} \! \int \! \! d^4x_i  \right) \frac{\delta^n S[v]}{\delta \phi(x_1) \cdots \delta\phi(x_n)} \varphi(x_1) \cdots \varphi(x_n),
\end{equation}
where $\varphi(x)=\phi(x)-v(x)$ is a shifted field.
Here, a two-particle irreducible diagram is a diagram which can not be cut in two by cutting only two internal lines, otherwise it is two-particle reducible. Various approximations can be made by truncating the diagrammatic expansion for $\Gamma_2[v,G]$. The mean field and gap equations are given as a stationary condition for $\Gamma[v,G]$ with respect to $v$ and $G$. From these equations, we can solve $G$ as a function of $v$, $G=G[v]$. Then the standard 1PI effective action is obtained by inserting $G[v]$ into $\Gamma[v,G]$, giving $\Gamma_{\mathrm{1PI}}[v]=\Gamma[v,G[v]]$.

\begin{figure}
 \includegraphics[width=10cm,clip]{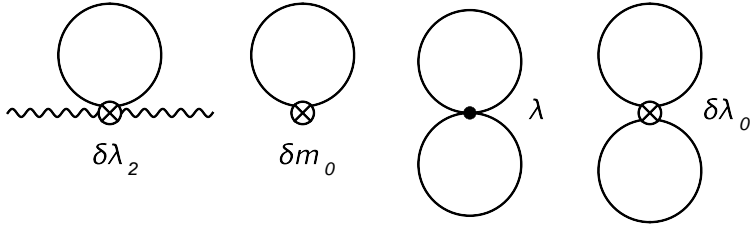}
 \caption{\label{fig:bubble} 2PI diagrams at the Hartree truncation level of the 2PI effective action. The wiggly line represents the vacuum expectation value of the quantum field, $v$.}
 \end{figure}

For $\phi^4$ theory with the following action
\begin{equation}
S[\phi]=-\! \int \! \! d^4x \Bigl[ \frac{1}{2} \phi (\Box +m^2+\delta m_2) \phi+\frac{1}{4!} (\lambda+\delta \lambda_4) \phi^4 \Bigr],
\end{equation}
$G_0^{-1}$ and $S_{\mathrm{int}}$ are respectively given by
\begin{equation}
i G_0^{-1}[v](x,x')=-\Bigl[ \Box+m^2+\frac{1}{2} \lambda v^2 \Bigr] \delta (x-x'),
\label{eq:g0}
\end{equation}
\begin{equation}
S_{\mathrm{int}}[\varphi]=-\! \int \! \! d^4x \biggl[ \frac{1}{3!}v \varphi^3+\frac{1}{4!} \varphi^4 \biggr] -\frac{1}{2}\! \int \! \! d^4x(\delta m_0+\frac{1}{2} \delta \lambda_2 v^2).
\end{equation}
In our convention, there are no counterterms in the definition of $G_0^{-1}$. Furthermore, the diagrams which are constructed from counterterm vertices with only one internal line are considered to be 2PI diagrams as shown in Fig. \ref{fig:bubble}. The contributions of these diagrams in the 2PI effective action are also represented as follows
\begin{equation}
(-i)^2 \frac{1}{2} \! \int \! \! d^4 x (\delta m_0+\frac{1}{2} \delta \lambda_2 v^2) G(x,x)=-\frac{1}{2} \! \int \! \! d^4 x \! \int \! \! d^4 x' (\delta m_0+\frac{1}{2} \delta \lambda_2 v^2) \delta (x-x') G(x',x).
\end{equation}
Therefore, comparing this expression to Eq. (\ref{eq:g0}), we find that one can effectively treat these diagrams as independent counterterms in $G_0^{-1}$. These counterterms do not necessarily coincide with those coming from the bare parameter in $S[v]$. This is a crucial point in our renormalization prescription.

In this paper, we approximate the theory by only including the double bubble diagram and corresponding counterterm diagrams needed at this approximation order as shown in Fig. \ref{fig:bubble}. This truncation corresponds to the Hartree-Fock approximation. In this case, the 2PI effective action is given by
\begin{equation}
\begin{split}
\Gamma [v,G]=&-\! \int \! \! d^4 x \Bigl[ \frac{1}{2} v(\Box+m^2+\delta m_2) v+\frac{1}{4!} (\lambda +\delta \lambda_4) v^4 \Bigr]+\frac{i}{2} \log \mathrm{det} [G^{-1}] \\
&-\frac{1}{2} \! \int \! \! d^4 x \bigl[ \Box +m^2+\delta m_0+\frac{1}{2} (\lambda +\delta \lambda_2) v^2 \bigr] G(x,x)-\! \int \! \! d^4 x \frac{1}{8} (\lambda +\delta \lambda_0) G^2(x,x).
\end{split}
\end{equation}
Note that one can assign the independent counterterms since we can freely select the contributions of the 2PI diagrams at a given truncation. The indices of the different counterterms refer to the power of $v$.

\section{2PI renormalization scheme in flat space}
In this section, we explicitly construct counterterms at the Hartree-Fock truncation of the 2PI effective action. First, we derive the equations of motion for $v$ and $G$, and construct counterterms to renormalize them. Then we show that the effective action is renormalized by the same counterterms.

\subsection{Renormalization of the equations of motion in flat space}
The equations of motion for $v$ and $G$ are given by varying $\Gamma[v,G]$ with respect to $v$ and $G$:
\begin{align}
&\Bigl[ \Box+m^2+\delta m_0+\frac{1}{2} (\lambda+\delta \lambda_2)v^2+\frac{1}{2} (\lambda+\delta \lambda_0) G(x,x)\Bigr] G(x,x')=-i \delta (x-x'), \\
-&\Bigl[\Box+m^2+\delta m_2+\frac{1}{6} (\lambda+\delta \lambda_4) v^2+\frac{1}{2} (\lambda+\delta \lambda_2) G(x,x)\Bigr] v(x)=0.
\end{align}
From these equations, we find that we require $\delta \lambda_0=\delta \lambda_2$ and $\delta m_0=\delta m_2 \equiv \delta m$ for a consistent renormalization. We also assume that $v$ is a constant because of the Poincare invariance of the vacuum state. As a consequence, we have very simple equations:
\begin{equation}
\Bigl[ \Box+m_{ph}^2 \Bigr]G(x,x')=-i \delta (x-x'),
\end{equation}
\begin{equation}
-\Bigl[ m_{ph}^2-\frac{1}{3} \lambda v^2+\frac{1}{6} (\delta \lambda_4-3 \delta \lambda_2)v^2 \Bigr] v=0.
\end{equation}
Here, the equation of motion for the propagator is the same as that of the free field, and we identified the physical mass $m_{ph}^2$ from the equation of motion for the propagator as follows
\begin{equation}
m_{ph}^2=m^2+\delta m+\frac{1}{2} (\lambda+\delta \lambda_2) v^2+\frac{1}{2} (\lambda+\delta \lambda_2)G(x,x).
\end{equation}
Again, one must require $\delta \lambda_4=3 \delta \lambda_2$ for a consistent renormalization of the equations of motion. This fact means that one needs only one-third of the coupling counterterm coming from the bare parameter $\lambda_B=\lambda+\delta \lambda_4$ as $\delta \lambda_2$ for a consistent renormalization in the Hartree-Fock approximation.

We now renormalize this mass equation. The key point is to explicitly know the divergence structure of the radiative corrections. In the dimensional regularization scheme, it is well known that $G(x,x)$ is expressed as follows
\begin{equation}
\begin{split}
G(x,x)&=\frac{m^2}{16 \pi^2} \Bigl[-\frac{2}{\epsilon}-1+\gamma+\log \frac{m^2}{4 \pi}+\mathcal{O}(\epsilon) \Bigr], \\
           &\equiv m^2 T_d+T_F(m^2),
\end{split}
\end{equation}
where $T_d=-\frac{1}{16\pi^2}\frac{2}{\epsilon}$ is a tadpole divergent term, $\epsilon=4-d$ is a regularization parameter, $d$ is the dimensionality of spacetime and $T_F$ expresses finite tadpole corrections. Inserting this expression into the mass equation, we obtain
\begin{equation}
m_{ph}^2=m^2+\delta m+\frac{1}{2} (\lambda+\delta \lambda_2) v^2+\frac{1}{2} (\lambda+\delta \lambda_2) \Bigl[ m_{ph}^2 T_d+T_F(m_{ph}^2) \Bigr].
\end{equation}
We can renormalize this equation using the MS-like scheme, that is, we can drop only the divergent terms by using the counterterms. This prescription leads to the following expression for the physical mass
\begin{equation}
m_{ph}^2=m^2+\frac{1}{2} \lambda v^2+\frac{1}{2} \lambda T_F(m_{ph}^2).
\end{equation}
We find that the counterterms must satisfy the following equation
\begin{equation}
\delta m+\frac{1}{2}\delta \lambda_2 v^2+\frac{1}{2} (\lambda+\delta \lambda_2) m_{ph}^2 T_d+\frac{1}{2} \delta \lambda_2 T_F=0.
\end{equation}
To explicitly construct the counterterms $\delta m$ and $\delta \lambda$, the central step is to use the renormalized expression for $m_{ph}^2$. In fact, plugging the expression $m_{ph}^2$ into the counterterm equation, we obtain
\begin{equation}
\delta m+\frac{1}{2}\delta \lambda_2 v^2+\frac{1}{2} (\lambda+\delta \lambda_2) \Bigl[ m^2+\frac{1}{2} \lambda v^2+\frac{1}{2} \lambda T_F \Bigr] T_d+\frac{1}{2} \delta \lambda_2 T_F=0.
\end{equation}

Next, we take the one-step renormalization~\cite{Fejos}. We are able to distinguish between the overall-divergence and the sub-divergences in this equation. The sub-divergences are the divergences caused by the divergent sub-diagrams. The nonperturbative counterterms are deduced from the conditions for the cancellation of the overall- and the sub-divergences. We assume that the terms proportional to $T_F$ represent the sub-divergences since $T_F$ is a tadpole correction. Then we require that the expressions for the overall-divergence and the sub-divergences independently vanish:
\begin{equation}
\delta m+\frac{1}{2} \lambda T_d m^2+\frac{1}{2} \delta \lambda_2 T_d m^2
+v^2 \biggl[ \frac{1}{2} (\lambda+\delta \lambda_2) \frac{1}{2} \lambda T_d+\frac{1}{2}\delta \lambda_2 \biggr]=0,
\end{equation}
\begin{equation}
T_F \biggl[ \frac{1}{2} (\lambda+\delta \lambda_2) \frac{1}{2} \lambda T_d+\frac{1}{2}\delta \lambda_2 \biggr]=0.
\end{equation}
Note that the divergent terms which are proportional to $v^2$ in the equation of the overall-divergence vanish by using the equation of the sub-divergences. 

The equation of the sub-divergences determines the coupling constant counterterm $\delta \lambda$:
\begin{equation}
\begin{split}
\delta \lambda_2&=-\frac{1}{2} \lambda^2 T_d \Bigl(1+\frac{1}{2} \lambda T_d \Bigr)^{-1}, \\
&=\lambda \sum_{n=1}^{\infty} \Bigl( -\frac{1}{2} \lambda T_d \Bigr)^n.
\end{split}
\end{equation}
Note that $\delta \lambda$ has infinite series of divergent terms. This fact justifies our renormalization prescription because it is anticipated by the BPHZ renormalization scheme in standard perturbation theory. On the other hand, the equation of the overall-divergence determines the mass counterterm $\delta m$:
\begin{equation}
\begin{split}
\delta m&=-\frac{1}{2} m^2 T_d (\lambda+\delta \lambda_2), \\
&=m^2 \sum_{n=1}^{\infty} \Bigl( -\frac{1}{2} \lambda T_d \Bigr)^n.
\end{split}
\end{equation}
Again, $\delta m$ has infinite series of divergent terms. Note also that these counterterms can be obtained using the iterative procedure~\cite{Blaizot2}. With the aid of these counterterms, the equations of motion are properly renormalized to
\begin{equation}
\Bigl[ \Box+m_{ph}^2 \Bigr]G(x,x')=-i \delta (x-x'),
\end{equation}
\begin{equation}
\Bigl[ m_{ph}^2-\frac{1}{3} \lambda v^2 \Bigr] v=0.
\end{equation}

\subsection{Renormalization of the effective action in flat space}
We can now renormalize the effective action by using the counterterms obtained in the previous subsection. The 2PI effective action reads
\begin{equation}
\begin{split}
\Gamma [v,G]=&-\! \int \! \! d^4 x \Bigl[ \frac{1}{2} (m^2+\delta m) v^2+\frac{1}{4!} (\lambda +\delta \lambda_4) v^4 \Bigr]+\frac{i}{2} \log \mathrm{det} [G^{-1}] \\
&-\frac{1}{2} \! \int \! \! d^4 x \Bigl[ \Box +m^2+\delta m+\frac{1}{2} (\lambda +\delta \lambda_2) v^2 \Bigr] G(x,x)-\! \int \! \! d^4 x \frac{1}{8} (\lambda +\delta \lambda_0) G^2(x,x).
\end{split}
\end{equation}
We can eliminate the kinetic term for $G$ by using the equation of motion for $G$ in the 2PI effective action:
\begin{equation}
\begin{split}
\Gamma [v,G]=&-\! \int \! \! d^4 x \Bigl[ \frac{1}{2} (m^2+\delta m) v^2+\frac{1}{4!} (\lambda +\delta \lambda_4) v^4 \Bigr] \\
&-\frac{1}{2} \! \int \! \! d^4 x \! \int \! \! dm_{ph}^2 G(x,x)
                                      +\! \int \! \! d^4 x \frac{1}{8} (\lambda +\delta \lambda_2) G^2(x,x),
\end{split}
\label{eq:effective action in flat space}
\end{equation}
where we use the relation for the free field propagator and the one-loop effective action:
\begin{equation}
\Gamma_{\mbox{{\scriptsize 1-loop}}}=-\frac{1}{2} \! \int \! \! d^4 x \! \int \! \! dm^2 G(x,x).
\end{equation}
Now we explicitly calculate the second and third terms in Eq. (\ref{eq:effective action in flat space}). First, the second term is
\begin{equation}
\begin{split}
\! \int \! \! dm_{ph}^2 G(x,x)&=\! \int \! \! dm_{ph}^2 (m_{ph}^2 T_d+T_F), \\
&=\frac{1}{2} m^4 T_d+v^2 \biggl[ \frac{1}{2} m^2 \lambda T_d \biggr]+v^4 \biggl[ \frac{1}{8} \lambda^2 T_d \biggr]+v^2 T_F \biggl[ \frac{1}{4} \lambda^2 T_d \biggr] \\
&\ \ \ +T_F \biggl[ \frac{1}{2} m^2 \lambda T_d \biggr]+T_F ^2 \biggl[ \frac{1}{8} \lambda^2 T_d \biggr]+\! \int \! \! dm_{ph}^2 T_F.
\end{split}
\end{equation}
The third term is
\begin{equation}
\begin{split}
G^2 (x,x)=&\Bigl ( (m^2+\frac{1}{2}\lambda v^2+\frac{1}{2} \lambda T_F ) T_d+T_F \Bigr)^2, \\
=&m^4 T_d^2+v^2 \biggl[ m^2 \lambda T_d^2 \biggr]+v^4 \biggl[ \frac{1}{4} \lambda^2 T_d^2 \biggr]+v^2 T_F \biggl[ \lambda T_d (1+\frac{1}{2} \lambda T_d) \biggr] \\
&+T_F \biggl[ 2 m^2 T_d (1+\frac{1}{2} \lambda T_d) \biggr]+T_F^2 \biggl[ (1+\frac{1}{2} \lambda T_d)^2 \biggr].
\end{split}
\end{equation}
With the aid of these expressions, we are able to show that the divergent terms which are proportional to $v^2$, $v^4$, $v^2 T_F$, $T_F$ and $T_F^2$ in the 2PI effective action independently vanish by using the expressions for $(\lambda +\delta \lambda_2)=\lambda (1+\lambda T_d/2)^{-1}$ and $\delta m$. Then the renormalized 2PI effective action reads
\begin{equation}
\Gamma [v,G]=\! \int \! \! d^4 x \Bigl[ -\frac{1}{2} m^2 v^2-\frac{1}{24} \lambda v^4-\frac{1}{4} m^4 T_d+\frac{1}{8} (\lambda+\delta \lambda_2) m^4 T_d^2+\frac{1}{8} \lambda T_F^2-\frac{1}{2} \! \int \! \! dm_{ph}^2 T_F \Bigr].
\end{equation}
Removing the physically irrelevant divergent terms, we finally obtain the following expression for the renormalized 2PI effective action
\begin{equation}
\Gamma [v,G]=\! \int \! \! d^4 x \Bigl[ -\frac{1}{2} m^2 v^2-\frac{1}{24} \lambda v^4+\frac{1}{8} \lambda T_F^2-\frac{1}{2} \! \int \! \! dm_{ph}^2 T_F \Bigr].
\end{equation}

\section{2PI renormalization scheme in de Sitter space}
In this section, we extend our previous renormalization prescription to de Sitter space. We use the coordinate system for de Sitter space in terms of comoving spatial coordinates $\mathbf{x}$ and conformal time $-\infty < \eta < 0$ in which the metric takes the form
\begin{equation}
\begin{split}
ds^2&=dt^2-e^{2 H t} d\mathbf{x}^2, \\
         &=a(\eta)^2 (d\eta^2-d\mathbf{x}^2),
\end{split}
\end{equation}
where $a(\eta)=-1/H\eta$ is a scale factor and $H$ is a Hubble parameter constant.
For this geometry, the matter action for $\phi^4$ scalar fields reads
\begin{equation}
S_m[\phi,g^{\mu \nu}]=-\! \int \! \! d^4x \sqrt{-g} \Bigl[ \frac{1}{2} \phi (\Box +m^2+\delta m_2+\xi R +\delta \xi_2 R) \phi+\frac{1}{4!} (\lambda+\delta \lambda_4) \phi^4 \Bigr],
\end{equation}
where $\Box=g^{\mu \nu} \nabla_{\mu} \nabla_{\nu}$, $\nabla_{\mu}$ is a covariant derivative, $R=d (d-1)H^2$ is the Ricci scalar curvature and $\xi$ is the conformal factor, the coupling constant to gravity which is necesssary for the field theory to be renormalizable.

In this coordinate system for de Sitter space, the metric has a time dependence and its nonequilibrium nature may appear. In such a situation, it is known that the standard in-out formalism is not sufficient and it is more appropriate to take the Schwinger-Keldysh formalism~\cite{Ramsey}. In this paper however, we omit the closed-time path index for the Schwinger-Keldysh formalism since for our approximation order, these in-in and in-out formalisms give the same results.

In the realm of quantum field theory in curved spacetime, it is well known that one must add the following bare gravitational action with higher derivative terms to properly renormalize the matter effective action
\begin{equation}
S_g[g^{\mu \nu}]=\frac{1}{16 \pi G_B} \! \int \! \! d^4 x \sqrt{-g} \bigl( R-2 \Lambda_B+c_B R^2+b_B R^{\mu \nu} R_{\mu \nu}+a_B R^{\mu \nu \rho \sigma}R_{\mu \nu \rho \sigma} \bigr),
\end{equation}
where $R_{\mu \nu}$ is the Ricci tensor, $R_{\mu \nu \rho \sigma}$ is the Riemann tensor  and $\Lambda$ is the cosmological constant. The index $B$ means that they are understood to be bare. As a result of the generalized Gauss-Bonnet theorem, the constants $a_B$, $b_B$ and $c_B$ are not all independent in four spacetime dimensions~\cite{Ramsey}; let us, therefore, set $a_B$ to zero. 

In curved spacetime, the 2PI effective action of the matter field is modified as follows
\begin{equation}
\begin{split}
\Gamma[v,G,g^{\mu \nu}]=&-\! \int \! \! d^4 x \sqrt{-g} \Bigl[ \frac{1}{2} \phi (\Box +m^2+\delta m_2+\xi R+\delta \xi_2 R) \phi+\frac{1}{4!} (\lambda+\delta \lambda_4) \phi^4 \Bigr] \\
&-\frac{1}{2} \! \int \! \! d^4 x \sqrt{-g} [\Box+m^2+\delta m_0+\xi R+\delta \xi_0 R+\frac{1}{2} (\lambda+\delta \lambda_2) v^2] G(x,x) \\
&+\frac{i}{2} \log \mathrm{det} [G^{-1}] -\! \int \! \! d^4 x \sqrt{-g} \frac{1}{8} (\lambda+\delta \lambda_0) G^2(x,x).
\end{split}
\end{equation}

\subsection{Renormalization of the equations of motion in de Sitter space}
After the example of the renormalization prescription in flat space, we first renormalize the mean field and the gap equations in de Sitter space. The renormalization prescription proceeds in a similar way to the case of flat space.

Varying the matter effective action with respect to $v$ and $G$, we obtain the following equations of motion
\begin{equation}
\begin{split}
\sqrt{-g} \Bigl[ \Box+m^2+\delta m_0+(\xi+\delta \xi_0) R+\frac{1}{2} (\lambda+\delta \lambda_2) v^2+\frac{1}{2} (\lambda+\delta \lambda_0) G(x,x) \Bigr] G(x,y)& \\
=-i \delta (x-y)&,
\end{split}
\end{equation}
\begin{equation}
-\sqrt{-g}\Bigl[ \Box+m^2+\delta m_2+(\xi+\delta \xi_2) R+\frac{1}{6} (\lambda+\delta \lambda_4) v^2+\frac{1}{2} (\lambda+\delta \lambda_2) G(x,x) \Bigr] v(x)=0.
\end{equation}
As in the case of flat space, these equations are renormalized by resorting $\delta \lambda_0=\delta \lambda_2$, $\delta m_0=\delta m_2 \equiv \delta m$ and $\delta \xi_0=\delta \xi_2 \equiv \delta \xi$. We also assume that $v$ is a constant due to the de Sitter invariance of the vacuum state. Once again, one must require $\delta \lambda_4=3 \delta \lambda_2$ for a consistent renormalization. The equations of motion then read
\begin{align}
&\sqrt{-g} [\Box+m_{ph}^2+\xi R] G(x,x')=-i \delta (x-x'), \\
&[m_{ph}^2+\xi R-\frac{1}{3} \lambda v^2] v=0.
\end{align}
Here we identified the physical mass $m_{ph}^2$ from the equation of motion for the propagator as follows
\begin{equation}
m_{ph}^2+\xi R=m^2+\delta m+(\xi+\delta \xi) R+\frac{1}{2} (\lambda+\delta \lambda_2) v^2+\frac{1}{2} (\lambda+\delta \lambda_2) G(x,x).
\end{equation}
The one-step renormalization procedure proceeds in a similar way to the case of flat space.

The coincident propagator in de Sitter space is generally expressed as follows (see Appendix)
\begin{equation}
G(x,x)=(m^2+\kappa H^2) T_d+T_F(m^2),
\end{equation}
where $T_d$ is the tadpole divergent terms and $T_F$ is the finite tadpole corrections. Plugging this expression into the mass equation, we obtain
\begin{equation}
m_{ph}^2+\xi R=m^2+\delta m+(\xi+\delta \xi)R+\frac{1}{2} (\lambda+\delta \lambda_2) v^2+\frac{1}{2} (\lambda+\delta \lambda_2) \Bigl[ (m_{ph}^2+\kappa H^2) T_d+T_F(m_{ph}^2) \Bigr].
\end{equation}
Again, we renormalize this equation using the MS-like scheme and we only drop the divergent terms by using  the counterterms. This prescription leads to the following expression for the physical mass
\begin{equation}
m_{ph}^2=m^2+\frac{1}{2} \lambda v^2+\frac{1}{2} \lambda T_F.
\label{eq:mass}
\end{equation}
Then the counterterms have to satisfy
\begin{equation}
\delta m+\delta \xi R+\frac{1}{2} \delta \lambda_2 v^2+\frac{1}{2} (\lambda+\delta \lambda_2) (m_{ph}^2+\kappa H^2) T_d+\frac{1}{2} \delta \lambda_2 T_F=0.
\end{equation}
The central step for the renormalization is to use the renormalized expression for $m_{ph}^2$. Again, we also assume that the terms which depend on $T_F$ represent the sub-divergences, and that the overall-divergence and the sub-divergences independently vanish:
\begin{equation}
\delta m+\delta \xi R+\frac{1}{2} (\lambda+\delta \lambda_2) (m^2+\kappa H^2) T_d+v^2 \biggl[ \frac{1}{2} \delta \lambda_2+\frac{1}{2} (\lambda+\delta \lambda_2) \frac{1}{2} \lambda T_d \biggr]=0,
\end{equation}
\begin{equation}
T_F \biggl[ \frac{1}{2} \delta \lambda_2+\frac{1}{2} (\lambda+\delta \lambda_2) \frac{1}{2} \lambda T_d \biggr]=0.
\end{equation}
Note that the divergent terms proportional to $v^2$ in the equation of the overall-divergence vanish by using the equation of the sub-divergences. Moreover, the equation of the sub-divergences is exactly the same as for the case of flat space. This fact means that in de Sitter space, the coupling counterterm has the same value as in the case of flat space in our renormalization scheme. The equation of the overall-divergence determines the mass counterterm $\delta m$ and the conformal counterterm $\delta \xi$:
\begin{equation}
\Bigl[
\delta m+\frac{1}{2} \lambda m^2 T_d (\lambda+\delta \lambda_2)
\Bigr]
+\Bigl[
\delta \xi R+\frac{1}{2} \kappa H^2 T_d (\lambda+\delta \lambda_2)
\Bigr]=0.
\label{eq:counterterm in de Sitter space}
\end{equation}
Again, the first term is the same as for the case of flat space. Here, we assume that the first term in Eq. (\ref{eq:counterterm in de Sitter space}) vanishes by using the mass counterterm which has the same expression as for flat space. The residual divergences are renormalized by the conformal counterterm:
\begin{equation}
\begin{split}
\delta \xi R&=-\kappa H^2 \frac{1}{2} \lambda T_d (\lambda+\delta \lambda_2), \\
                   &=-\kappa H^2 \sum_{n=1}^{\infty} \Bigl( -\frac{1}{2} \lambda T_d \Bigr)^n.
\end{split}
\end{equation}
Again, the conformal counterterm has infinite series of divergent terms. 

Note that the counterterms, $\delta m$ and $\delta\lambda$, coincide with those of flat space and they have no geometrical dependences. All the divergences that depend on the geometrical parameter in de Sitter space are renormalized with the conformal counterterm.

\subsection{Renormalization of the effective action in de Sitter space}
Next, we renormalize the effective action. In contrast to the flat space case, it is well known that the nature of curved spacetime gives further divergences which can only be renormalized with the gravitational counterterms, the redefinition of coupling constants in the gravitational action. That is, $\Gamma[v,G,g^{\mu \nu}]$ cannot be finite by itself, but the sum $S_g+\Gamma$ can be finite. To see this, we first express the divergence structure of the 2PI effective action of the matter field. As in the case of flat space, the 2PI effective action can be transformed to
\begin{equation}
\begin{split}
\Gamma[v,G,g^{\mu \nu}]=&-\! \int \! \! d^4 x \sqrt{-g} \Bigl[ \frac{1}{2} (m^2+\delta m+(\xi+\delta \xi )R)v^2+\frac{1}{4!} (\lambda+\delta \lambda_4) v^4 \Bigr] \\
&-\frac{1}{2} \! \int \! \! d^4 x \sqrt{-g} \! \int \! \! dm_{ph}^2 G(x,x)+\! \int \! \! d^4 x \sqrt{-g} \frac{1}{8} (\lambda+\delta \lambda_2) G^2 (x,x).
\end{split}
\label{eq:matter effective action}
\end{equation}
Again, we explicitly calculate the second and third terms in Eq. (\ref{eq:matter effective action}). The second term is 
\begin{equation}
\begin{split}
\! \int \! \! dm_{ph}^2 G(x,x)&=\! \int \! \! dm_{ph}^2 \Bigl[ m_{ph}^2 T_d+\kappa H^2 T_d+T_F \Bigr], \\
                                            &=\frac{1}{2} m_{ph}^4 T_d+\kappa H^2 m_{ph}^2 T_d+\! \int \! \! dm_{ph}^2 T_F, \\
                                            &=(m^2+\xi R) \kappa H^2 T_d+\frac{1}{2} (m^2 +\xi R)^2 T_d+ v^2 \biggl[ \frac{1}{2} \lambda T_d (m^2+\kappa H^2) \biggr]
                                            +v^4 \biggl[ \frac{1}{8} \lambda^2 T_d \biggr] \\ 
                                            &\ \ \ +v^2 T_F \biggl[ \frac{1}{4} \lambda^2 T_d \biggr]
                                            +T_F \biggl[ \frac{\lambda}{2} (m^2+\kappa H^2) \biggr]+T_F^2 \biggl[ \frac{1}{8} \lambda^2 T_d \biggr] +\! \int \! \! dm_{ph}^2 T_F.
\end{split}
\end{equation}
The third term is
\begin{equation}
\begin{split}
G^2 (x,x)&=\Bigl[ (m_{ph}^2+\kappa H^2)T_d+T_F \Bigr]^2, \\
                &=\bigl( m^2+\kappa H^2 \bigr)^2 T_d^2+v^2 \biggl[ \lambda T_d^2 (m^2+\kappa H^2) \biggr]+v^4 \biggl[ \frac{1}{4} \lambda^2 T_d^2\biggr] \\
                &\ \ \ +v^2 T_F \biggl[ \lambda T_d (1+\frac{1}{2} \lambda T_d)\biggr] +T_F \biggl[ 2 (m^2+\kappa H^2) T_d (1+\frac{1}{2} \lambda T_d) \biggr]+T_F^2 \biggl[ (1+\frac{1}{2} \lambda T_d)^2\biggr].
\end{split}
\end{equation}
As in the case of flat space, we can show that the divergent terms which depend on $v^2$, $v^4$, $v^2T_F$, $T_F$ and $T_F^2$ in the 2PI effective action independently vanish by using the expressions for $(\lambda +\delta \lambda_2)=\lambda (1+\lambda T_d/2)^{-1}$, $\delta m$ and $\delta \xi$. Finally, the 2PI effective action is given by
\begin{equation}
\begin{split}
\Gamma[v,G,g^{\mu \nu}]=&-\! \int \! \! d^4 x \sqrt{-g} \Bigl[\frac{1}{2} (m^2+\xi R) v^2+\frac{1}{24} \lambda v^4-\frac{1}{8} \lambda T_F^2+\frac{1}{2}\! \int \! \! dm_{ph}^2 T_F  \Bigr] \\
 & +\! \int \! \! d^4 x \sqrt{-g} \Bigl[-\frac{1}{2} (m^2+\xi R) \kappa H^2 T_d-\frac{1}{4} (m^2+\xi R)^2 T_d \\
 &\hspace{6.2cm}+\frac{1}{8} (\lambda +\delta \lambda_2) \bigl( m^2+\kappa H^2 \bigr)^2T_d^2 \Bigr].
\end{split}
\label{eq:action}
\end{equation}
The last terms in this expression are divergent and are not renormalized by the effective action of the matter field itself. Thus one must resort to the redefinition of the coupling constants in the gravitational action. To this aim, we re-express the $H$ dependent divergent terms $\kappa H^2$ as purely geometrical expressions. In the case of the minimal coupling, this term is expressed by the Ricci scalar curvature: $\kappa H^2=-2 H^2=-2 R/d (d-1)\equiv \zeta R$. In the case of the conformal coupling, $\kappa H^2\equiv \zeta R=0$. Then the last terms in Eq. (\ref{eq:action}) are expressed as follows
\begin{equation}
\begin{split}
\Gamma_{\mathrm{div}}&\equiv -\frac{1}{2} (m^2+\xi R) \kappa H^2 T_d-\frac{1}{4} (m^2+\xi R)^2 T_d+\frac{1}{8} (\lambda +\delta \lambda_2) \bigl( m^2+\kappa H^2 \bigr)^2T_d^2,
 \\
&=-\frac{1}{2} (m^2+\xi R) \zeta R T_d-\frac{1}{4} (m^2+\xi R)^2 T_d+\frac{1}{8} (\lambda +\delta \lambda_2) \bigl( m^2+\zeta R \bigr)^2T_d^2,
 \\
&=-\frac{1}{4} m^4 T_d (1-\frac{1}{2} (\lambda+\delta \lambda_2)T_d) \\
&\ \ \ -\frac{1}{2} T_d R m^2 \biggl\{ \xi+\zeta \Bigl[1-\frac{1}{2} (\lambda+\delta \lambda_2)T_d \Bigr] \biggr\}
-\frac{1}{4} R^2 T_d \biggl\{ \xi^2+\zeta^2 \Bigl[1-\frac{1}{2} (\lambda+\delta \lambda_2)T_d \Bigr] \biggr\}.
\end{split}
\label{eq:div}
\end{equation}
Note that in the minimally coupled field, $\Gamma_{\mathrm{div}}$ has only divergent terms, and has no finite terms. In Eq. (\ref{eq:div}), the first term is renormalized by $\Lambda_B$, the second term is renormalized by the $R$ term , and the third term is renormalized by the $R^2$ term in the gravitational action. That is, these divergent terms are renormalized using the following redefinitions of the coupling constants in the gravitational action
\begin{equation}
\frac{1}{16 \pi G_B} (-2 \Lambda_B)-\frac{1}{4} m^4 T_d (1-\frac{1}{2}(\lambda+\delta \lambda_2)T_d)=\frac{1}{16 \pi G} (-2 \Lambda),
\end{equation}
\begin{equation}
\Biggl( \frac{1}{16 \pi G_B} -\frac{1}{2} m^2 T_d \biggl\{ \xi+\zeta \Bigl[1-\frac{1}{2} (\lambda+\delta \lambda_2)T_d \Bigr] \biggr\} \Biggr) R= \frac{1}{16 \pi G}R,
\end{equation}
\begin{equation}
\Biggl( \frac{1}{16 \pi G_B}c_B -\frac{1}{4} T_d \biggl\{ \xi^2+\zeta^2 \Bigl[1-\frac{1}{2} (\lambda+\delta \lambda_2) T_d \Bigr] \biggr\} \Biggr) R^2=\frac{1}{16 \pi G} c R^2.
\end{equation}
\begin{equation}
\frac{1}{16 \pi G_B} b_B R^{\mu \nu} R_{\mu \nu}=\frac{1}{16 \pi G} b R^{\mu \nu} R_{\mu \nu}.
\end{equation}
Note also that in our renormalization prescription, $G$ and $c$ have finite terms in addition to divergent terms in the conformally coupled case. Such ambiguities of finite terms are always present when we determine the renormalized coupling constants in the gravitational action. With the aid of this renormalization, we find that the renormalized expression for $S_g+\Gamma$ is
\begin{equation}
\begin{split}
S_g [g^{\mu \nu}]+\Gamma[v,G,g^{\mu \nu}]=&\frac{1}{16 \pi G} \! \int \! \! d^4 x \sqrt{-g} \bigl( R-2 \Lambda+a R^2+b R^{\mu \nu} R_{\mu \nu} \bigr)  \\
&-\! \int \! \! d^4 x \sqrt{ -g}\Bigl[\frac{1}{2} (m^2+\xi R) v^2+\frac{1}{24} \lambda v^4-\frac{1}{8} \lambda T_F^2+\frac{1}{2}\! \int \! \! dm_{ph}^2 T_F \Bigr], \\
\equiv&S_g^{\mathrm{ren}}[g^{\mu \nu}]+\Gamma^{\mathrm{ren}} [v,G,g^{\mu \nu}].
\end{split}                                  
\end{equation}
Einstein's field equation with the quantum matter backreaction is obtained as the stationary condition by differentiating the action $S_g^{\mathrm{ren}}+\Gamma^{\mathrm{ren}}$ with respect to the metric $-2 \delta/\bigl(\sqrt{-g} \delta g^{\mu \nu}\bigr)$. Note that the classical Einstein equation is reproduced only in the limited case $c=b=0$. More concretely, the vacuum expectation value of the energy-momentum tensor $T_{\mu \nu}\equiv -2\delta S_m/\bigl(\sqrt{-g} \delta g^{\mu \nu}\bigr)$ is obtained by
\begin{equation}
\langle T_{\mu \nu} \rangle \equiv \frac{\int D\phi T_{\mu \nu} e^{i S_m}}{\int D \phi e^{i S_m}} =\frac{-2}{\sqrt{-g}} \frac{\delta \Gamma^{\mathrm{ren}}}{\delta g^{\mu \nu}}=\Bigl(-\frac{1}{2} (m^2+\xi R) v^2-\frac{1}{24} \lambda v^4+\frac{1}{8} \lambda T_F^2-\frac{1}{2} \! \int \! \! dm_{ph}^2 T_F\Bigr) g_{\mu \nu}.
\end{equation}

\section{influence of the mass generation for minimally coupled fields}
In the previous section, we showed that we can consistently renormalize the effective action and the energy-momentum tensor on full de Sitter geometry at the Hartree truncation level of the 2PI effective action. In this section, we investigate the physical influence of the dynamical mass generation for the minimally coupled light fields using this renormalized effective action.

\subsection{Evaluation of the physical mass}
We first solve the mass equation as a function of $v$ in order to obtain the 1PI effective action from the 2PI effective action. The equation of the physical mass is given as in Eq. (\ref{eq:mass})
\begin{equation}
m_{ph}^2=m^2+\frac{1}{2} \lambda v^2+\frac{1}{2} \lambda T_F(m_{ph}^2).
\end{equation}
Using the lowest order expression of the small mass expansion of $T_F$ (see Appendix), we obtain the mass equation as an algebraic equation
\begin{equation}
m_{ph}^4-(m^2+\frac{1}{2} \lambda v^2) m_{ph}^2-\frac{3 \lambda H^4}{16 \pi^2}=0.
\end{equation}
The physically meaningful solution of this equation is given by
\begin{equation}
m_{ph}^2(v)=\frac{1}{2} \biggl \{m^2+\frac{1}{2} \lambda v^2+\sqrt{(m^2+\frac{1}{2}\lambda v^2)^2+\frac{3 \lambda H^4}{4 \pi^2}} \biggr \}.
\label{eq:physical mass}
\end{equation}
From this expression we see that $m_{ph}^2$ never vanishes. The physical mass always acquires a positive term due to the second term in the square root in Eq. (\ref{eq:physical mass}). That is, in the theory with a mass parameter $m^2/H^2 \ll \sqrt{3 \lambda} /2\pi$, the infrared divergence existing in the propagator is regulated by the dynamically generated mass term instead of the mass parameter $m$.

\subsection{Evaluation of the effective potential in a broken phase}
\begin{figure}
 \includegraphics[width=12cm,clip]{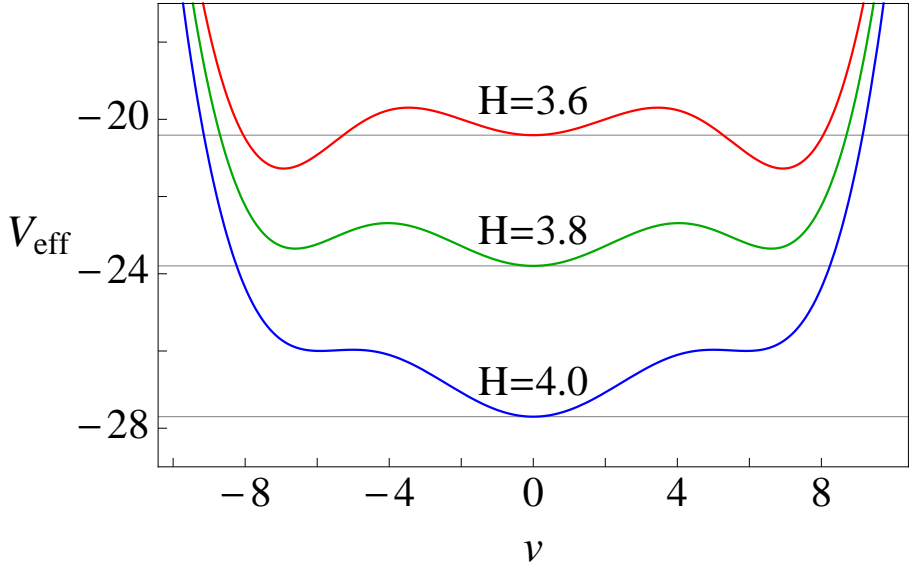}
 \caption{\label{fig:potential} The effective potentials as a function of $v$ for $\lambda=0.1$ all in the units of $|m|$. The different lines show the potentials with different values of $H$.}
 \end{figure}

Next, we evaluate the 2PI resummed effective potential for tachyonic mass parameters. The renormalized effective potential reads
\begin{equation}
V_{\mathrm{eff}}(v)=\frac{1}{2} m^2 v^2+\frac{1}{24} \lambda v^4-\frac{1}{8} \lambda T_F^2+\frac{1}{2}\! \int \! \! dm_{ph}^2 T_F.
\label{eq:effective potential}
\end{equation}
For generality of the discussion, we express the small mass expansion of the tadpole correction $T_F$ as follows
\begin{equation}
T_F=\frac{H^2}{16 \pi^2} \biggl( b_{-1} \frac{H^2}{m_{ph}^2}+b_0+b_1 \frac{m_{ph}^2}{H^2}+b_2 \Bigl( \frac{m_{ph}^2}{H^2} \Bigr)^2+\mathcal{O}((\tfrac{m_{ph}^2}{H^2})^3) \biggr).
\end{equation}
Then, the third and fourth terms in Eq. (\ref{eq:effective potential}) are calculated as follows
\begin{equation}
T_F^2=\Bigl(\frac{H^2}{16 \pi^2} \Bigr)^2  \biggl(b_{-1}^2 \Bigl( \frac{H^2}{m_{ph}^2} \Bigr)^2+2b_{-1} b_0 \frac{H^2}{m_{ph}^2}+b_0^2+2 b_{-1} b_1+\mathcal{O}(\tfrac{m_{ph}^2}{H^2}) \biggr),
\end{equation}
\begin{equation}
\int \! \! dm_{ph}^2 T_F=\frac{H^4}{16\pi^2} \biggl(b_{-1} \log \frac{m_{ph}^2}{H^2}+\mathcal{O}(\tfrac{m_{ph}^2}{H^2}) \biggr).
\end{equation}
If we take the lowest order expression of the small mass expansion for $T_F$, we obtain the following expression for the effective potential
\begin{equation}
V_{\mathrm{eff}}(v) \simeq \frac{1}{2} m^2 v^2+\frac{1}{24} \lambda v^4-\frac{9}{2}\lambda \Bigl(\frac{H^4}{16^2 \pi^4} \Bigr) \Bigl( \frac{H^4}{m_{ph}^4(v)} \Bigr)+ \frac{3H^4}{16 \pi^2} \log \frac{m_{ph}^2(v)}{H^2}.
\end{equation}
The behavior of the effective potential as a function of $v$ near the phase transition is displayed in Fig. \ref{fig:potential}. This result is consistent with our previous analysis which shows a first-order phase transition~\cite{Arai}. However, in contrast to the previous result which expresses only $v$ dependent contributions, we obtain the effective potential with $v$ independent constants. This result can never be obtained without the proper renormalization prescriptions.

\subsection{Renormalized energy-momentum tensor}
In this subsection, we take $m^2 >0$, that is, we are in the symmetric phase $v=0$, and investigate whether there are any differences in the energy-momentum tensor between vanishing renormalized mass parameters and massive light fields. The energy-momentum tensor is given by the functional differentiation of the renormalized effective action $\Gamma^{\mathrm{ren}}[v,g^{\mu \nu}]$ with respect to the metric tensor:
\begin{equation}
\begin{split}
\langle T_{\mu \nu} \rangle \simeq \left\{\frac{9}{2}\lambda \Bigl(\frac{H^4}{16^2 \pi^4} \Bigr) \Biggl[ \frac{m^2+\sqrt{m^4+\tfrac{3 \lambda H^4}{4 \pi^2}} }{4H^4} \Biggr]^{-2}
                                                   -\frac{3H^4}{16 \pi^2} \log \frac{m^2+\sqrt{m^4+\tfrac{3 \lambda H^4}{4 \pi^2}}}{2H^2} \right\}g_{\mu \nu}.
\end{split}
\end{equation}
Note that the last term is the one-loop contribution. In the massless limit we obtain
\begin{equation}
\langle T_{\mu \nu} \rangle=\biggl \{ \frac{3H^4}{32 \pi^2}-\frac{3H^4}{16\pi^2}\log \frac{\sqrt{3 \lambda}}{4 \pi}+\mathcal{O}(\sqrt{\lambda}) \biggr \} g_{\mu \nu}.
\label{eq:energy-momentum tensor}
\end{equation}
Of course, the proportionality of the energy-momentum tensor to the metric tensor is anticipated by the de Sitter symmetry. Note also that the first term in Eq. (\ref{eq:energy-momentum tensor}) depends on the renormalization condition for the gravitational counterterms.  Again, for the theory with the small mass parameter, $m^2/H^2 \ll \sqrt{3 \lambda}/2 \pi$, the infrared enhanced term in the energy-momentum tensor is regulated by the dynamically generated mass term instead of the mass parameter. We insist that this effect is a genuine nonperturbative effect in de Sitter space, and can be never captured by the perturbative expansion.

Moreover, the backreaction of the quantum matter field works on contracting the expanding universe, $\langle T_{\mu \nu} \rangle>0$,  when the coupling constant is small, $\lambda \ll1$ which is a necessary condition for our small mass expansion. For a vanishing renormalized mass parameter, this backreaction can be sufficiently large if we take a sufficiently small coupling constant $\lambda$.

\section{Conclusion}
In this paper, we elaborated our previous analysis, the analysis for light (including massless) scalar fields using the 2PI formalism at the Hartree truncation level, in the direction of the renormalization prescription. An MS-like scheme is possible, and due to its nonperturbative nature, one needs an infinite series of divergent terms as counterterms for a consistent renormalization. Investigating the divergence structure of a tadpole correction, we find that there are divergences which are the same as in the case of flat space as well as those that are specific to curved space. Divergences analogous to flat space are renormalized using the mass counterterm $\delta m$ and the coupling constant counterterm $\delta \lambda$, which are the same expressions as for flat space. Divergences inherent to curved space are renormalized by the conformal counterterm $\delta \xi$ and the redefinition of the coupling constants in the gravitational action. A divergence specific to curved space in the propagator vanishes for the conformally coupled case.

Using this renormalization prescription, we demonstrated the mass generation which is the same as the result found in the previous analysis~\cite{Arai}. The present renormalization scheme further enables us to calculate the phase structure and the vacuum expectation value of the energy-momentum tensor up to their absolute value. Note that in the other renormalization prescription, our present results can be never obtained. As a result of the mass generation, infrared enhanced terms which are present in the propagator and the energy-momentum tensor are regulated by the dynamically generated mass term. Otherwise the infrared enhanced terms can be indefinitely large if we take the mass parameter to be sufficiently small. These facts show that infrared divergences in the perturbative expansion in full de Sitter space arise due to the unsuitableness of the perturbative expansion around the massless minimally coupled free field, and can be circumvented by using the proper method of calculation.

Furthermore, it may be possible to insist on the following proposal on nontrivial renormalizability in curved space. If the model is renormalizable in flat space, it is also renormalizable in curved space, and there are divergent structures similar to flat space. These divergences are renormalized by the counterterms, such as $\delta m$ and $\delta \lambda$, which are the same expressions as for flat space. It is expected that these counterterms never depend on the geometrical parameters. Divergences specific to curved space are renormalized by the parameters in the Lagrangian which can exist only in the case of curved space, for example, the conformal factor $\xi$.

Moreover, this result refers to the renormalization at finite temperature field theory. Since the very-short-distance behavior of the theory is unaffected by finite temperature, the same divergent structure should exist at finite temperature as for that at zero temperature. These divergences should be renormalized by the same counterterms as in the case of zero temperature, and these counterterms should not depend on the temperature. We believe that our present renormalization prescription enables us to renormalize more complicated models at finite temperature which were considered previously to be non-renormalizable in the resummation scheme. Further study of renormalization for more complicated models both in curved space and at finite temperature is desired in order to check this.

\begin{acknowledgments}
I would like to thank H. Kanno for a careful reading of the manuscript. I also thank U. Reinosa and J. Serreau for informing me about some useful references on renormalization of the 2PI formalism. This work was supported by the Grant-in-Aid for Nagoya University Global COE Program, "Quest for Fundamental Principles in the Universe: from Particles to the Solar System and the Cosmos", from the Ministry of Education, Culture, Sports, Science and Technology of Japan.
\end{acknowledgments}

\appendix*
\section{Coincident propagator in de Sitter space}
In this appendix, we calculate the coincident propagator in de Sitter space to investigate the divergence structure of the tadpole diagram. In de Sitter space, a propagator for a free scalar field with mass $m$, conformal factor $\xi$ and the dimensionality of spacetime $d$ is expressed by the hypergeometric function~\cite{Candelas}
\begin{equation}
G(x,x')=\frac{H^{d-2}}{(4\pi)^{d/2}}\frac{\Gamma(\frac{d-1}{2}+\nu) \Gamma(\frac{d-1}{2}-\nu)}{\Gamma(\frac{d}{2})} {}_2\mathrm{F}_1\left[\tfrac{d-1}{2}+\nu,\tfrac{d-1}{2}-\nu,\tfrac{d}{2};1+\tfrac{y}{4}\right], 
\end{equation}
where $\nu=\bigl\{ [(d-1)/2 ]^2-(m^2+\xi R)/H^2 \bigr\}^{1/2}$, $R=d (d-1) H^2$ is the Ricci scalar curvature and $y(x,x')=\bigl[ (\eta-\eta')^2-|\mathbf{x}-\mathbf{x}'|^2 \bigr]/\eta \eta '$ is the de Sitter invariant length. In the coincident limit, $y=0$, the formula of the hypergeometric function, ${}_2 \mathrm{F}_1(a,b,c;1)=\Gamma(c)\Gamma(c-a-b)/\bigl[\Gamma(c-a)\Gamma(c-b)\bigr]$, leads to
\begin{equation}
\begin{split}
G(x,x)&=\frac{H^{d-2}}{(4\pi)^{d/2}} \Gamma(1-\tfrac{d}{2})\frac{\Gamma(\frac{d-1}{2}+\nu)\Gamma(\frac{d-1}{2}-\nu)}{\Gamma(\frac{1}{2}+\nu) \Gamma(\frac{1}{2}-\nu)},  \\
& \equiv \frac{H^{d-2}}{(4 \pi)^{d/2}} \Gamma(1-\tfrac{d}{2}) \Gamma(x,x).
\end{split}
\end{equation}
The first Gamma function has an ultraviolet divergent pole. The residual gamma function $\Gamma(x,x)$ determines a coefficient of the ultraviolet divergent pole.

\subsection{Minimally coupled case}
Let us consider the $\xi=0$ case. In this case, we can transform the expression $\Gamma(x,x)$ as follows
\begin{equation}
\begin{split}
\Gamma(x,x)
&=\frac{\Gamma(1+\frac{d-3}{2}+\nu) \Gamma(1+\frac{d-3}{2}-\nu)}
             {\Gamma(\frac{1}{2}+\nu) \Gamma(\frac{1}{2}-\nu)}, \\
&=\Bigl( \frac{d-3}{2}+\nu \Bigr) \Bigl( \frac{d-3}{2}-\nu \Bigr)
      \frac{\Gamma(\frac{d-3}{2}+\nu) \Gamma(\frac{d-3}{2}-\nu)}
             {\Gamma(\frac{1}{2}+\nu) \Gamma(\frac{1}{2}-\nu)}, \\
&=\biggl( \Bigl( \frac{d-3}{2} \Bigr)^2- \Bigl( \frac{d-1}{2} \Bigr)^2+\frac{m^2}{H^2} \biggr) \\
      & \ \ \ \ \frac{
            \Gamma(\frac{1}{2}+\nu) \bigl[1+\psi(\frac{1}{2}+\nu)(-\frac{\epsilon}{2})+\mathcal{O}(\epsilon^2) \bigr]
             \Gamma(\frac{1}{2}-\nu) \bigl[1+\psi(\frac{1}{2}-\nu)(-\frac{\epsilon}{2})+\mathcal{O}(\epsilon^2) \bigr]
            }
            {
            \Gamma(\frac{1}{2}+\nu) 
            \Gamma(\frac{1}{2}-\nu)
            },  \\
&=\Bigl( \frac{m^2}{H^2}-2+\epsilon \Bigr) \biggl[1-\Bigl(\frac{\epsilon}{2}\Bigr) \Bigl( \psi(\tfrac{1}{2}+\nu)+\psi(\tfrac{1}{2}-\nu) \Bigr)+\mathcal{O}(\epsilon^2) \biggr], \\
&=\Bigl(\frac{m^2}{H^2}-2 \Bigr) \biggl[1-\Bigl(\frac{\epsilon}{2}\Bigr) \Bigl( \psi(\tfrac{1}{2}+\nu)+\psi(\tfrac{1}{2}-\nu) \Bigr)+\mathcal{O}(\epsilon^2) \biggr]+\epsilon,
\end{split}
\end{equation}
where $\psi(x)$ is the digamma function, and we restrict our attention to four dimensional spacetime with a regularization parameter $\epsilon=4-d$. We expand $\nu$ in powers of $\epsilon$:
\begin{equation}
\nu=\frac{3}{2}-s+\mathcal{O}(\epsilon), \ \ \ \ \ \ 
s=\frac{3}{2}-\biggl[ \Bigl( \frac{3}{2} \Bigr)^2-\frac{m^2}{H^2} \biggr]^{1/2}.
\end{equation}
Then $\Gamma (x,x)$ is further transformed into
\begin{equation}
\begin{split}
\Gamma(x,x)&=\Bigl(\frac{m^2}{H^2}-2 \Bigr) \biggl[1-\Bigl(\frac{\epsilon}{2}\Bigr) \Bigl( \psi(2-s)+\psi(-1+s) \Bigr)+\mathcal{O}(\epsilon^2) \biggr]+\epsilon, \\
                       &=\Bigl(\frac{m^2}{H^2}-2 \Bigr) \biggl[1-\Bigl(\frac{\epsilon}{2}\Bigr) \Bigl( \psi(1+s)+\psi(1-s)-\frac{1}{s}+\frac{2}{1-s} \Bigr)+\mathcal{O}(\epsilon^2) \biggr]+\epsilon,
\end{split}
\end{equation}
where we use the formula for digamma function, $\psi(1+x)=\psi(x)+1/x$.
Therefore for the minimally coupled field, the coincident propagator is given by
\begin{equation}
\begin{split}
G(x,x)&=\frac{H^2}{16 \pi^2} \Bigl(1-\Bigl( \frac{\epsilon}{2} \Bigr) \log \frac{H^2}{4\pi}+\mathcal{O}(\epsilon^2) \Bigr) \Bigl( -\frac{2}{\epsilon}-1+\gamma+\mathcal{O}(\epsilon) \Bigr) \\
& \ \ \ \ \Biggl\{ \Bigl( \frac{m^2}{H^2}-2 \Bigr) \biggl[1-\Bigl(\frac{\epsilon}{2}\Bigr) \Bigl( \psi(1+s)+\psi(1-s)-\frac{1}{s}+\frac{2}{1-s} \Bigr)+\mathcal{O}(\epsilon^2) \biggr]+\epsilon \Biggr\}, \\
&=\frac{H^2}{16 \pi^2} \Bigl( -\frac{2}{\epsilon}-1+\gamma+\mathcal{O}(\epsilon) \Bigr) \\
&\ \ \ \ \Biggl\{ \Bigl( \frac{m^2}{H^2}-2 \Bigr) \biggl[1-\Bigl(\frac{\epsilon}{2}\Bigr) \Bigl( \psi(1+s)+\psi(1-s)-\frac{1}{s}+\frac{2}{1-s}+\log\frac{H^2}{4 \pi} \Bigr)\biggr]+\epsilon +\mathcal{O}(\epsilon^2) \Biggr\}, \\
&=\frac{H^2}{16 \pi^2}\Biggl\{ \Bigl(\frac{m^2}{H^2}-2 \Bigr) \biggl[ -\frac{2}{\epsilon}+  \psi(1+s)+\psi(1-s)-\frac{1}{s}-1+\frac{2}{1-s}+\Bigl(\gamma+\log\frac{H^2}{4 \pi} \Bigr) \biggr] \\
& \hspace{12.7cm} -2+\mathcal{O}(\epsilon) \Biggr\}, 
\end{split}
\end{equation}
where $\gamma$ is the Euler-Mascheroni constant.

If we expand $s$ around the massless case assuming $m^2/H^2 \ll1$, we obtain
\begin{equation}
\begin{split}
G(x,x)=&-\frac{1}{16 \pi^2} (m^2-2 H^2) \frac{2}{\epsilon}+\frac{1}{16 \pi^2} (m^2-2 H^2) \bigl( \gamma+\log\frac{H^2}{4 \pi} \bigr) \\
             &+\frac{H^2}{16 \pi^2} \biggl[ \frac{6 H^2}{m^2}+4\gamma-\frac{23}{3}-\bigl( 2\gamma+\frac{2}{27} \bigr) \frac{m^2}{H^2} \biggr]+\mathcal{O} \bigl(\epsilon, (\tfrac{m^2}{H^2})^2 \bigr).
\end{split}
\end{equation}
Note particular that the coefficients of the ultraviolet pole are only $(m^2-2 H^2)$.

\subsection{Conformally coupled case}
In the conformal coupling case, the conformal factor $\xi$ is $(d-2)/4(d-1)$, which is determined by the conformal transformation symmetry of the action. In this case, note that the dimensionality dependence in $\nu$ disappears:
$\nu=\bigl[(1/2)^2-m^2/H^2\bigr]^{1/2}$.

Then we can transform $\Gamma(x,x)$ into the form
\begin{equation}
\begin{split}
\Gamma(x,x)
&=\frac{\Gamma(\frac{3-\epsilon}{2}+\nu) \Gamma(\frac{3-\epsilon}{2}-\nu)}{\Gamma(\frac{1}{2}+\nu) \Gamma(\frac{1}{2}-\nu)}, \\
&=\frac{
            \Gamma(\frac{3}{2}+\nu) \bigl[ 1+\psi(\frac{3}{2}+\nu)(-\frac{\epsilon}{2})+\mathcal{O}(\epsilon^2) \bigr]
             \Gamma(\frac{3}{2}-\nu) \bigl[ 1+\psi(\frac{3}{2}-\nu)(-\frac{\epsilon}{2})+\mathcal{O}(\epsilon^2) \bigr]
            }
            {
            \Gamma(\frac{1}{2}+\nu) 
            \Gamma(\frac{1}{2}-\nu)
            },  \\
&=\Bigl( \frac{1}{2}+\nu \Bigr) \Bigl( \frac{1}{2}-\nu \Bigr) \biggl[1-\Bigl(\frac{\epsilon}{2}\Bigr) \Bigl( \psi(\tfrac{3}{2}+\nu)+\psi(\tfrac{3}{2}-\nu) \Bigr)+\mathcal{O}(\epsilon^2) \biggr], \\
&=\frac{1}{H^2} (m^2) \biggl[1-\Bigl(\frac{\epsilon}{2}\Bigr) \Bigl( \psi(\tfrac{3}{2}+\nu)+\psi(\tfrac{3}{2}-\nu) \Bigr)+\mathcal{O}(\epsilon^2) \biggr], \\
&=\frac{m^2}{H^2} \biggl[ 1-\Bigl(\frac{\epsilon}{2}\Bigr) \Bigl( \psi(2-s)+\psi(1+s) \Bigr)+\mathcal{O}(\epsilon^2) \biggr],
\end{split}
\end{equation}
where $s$ is defined by
\begin{equation}
\nu=\frac{1}{2}-s, \ \ \ \ \ \ 
s=\frac{1}{2}-\biggl[ \Bigl( \frac{1}{2} \Bigr)^2-\frac{m^2}{H^2} \biggr]^{1/2}.
\end{equation}
Therefore, in the conformally coupled case, the coincident propagator is given by
\begin{equation}
\begin{split}
G(x,x)&=\frac{H^2}{16 \pi^2} \Bigl(1- \bigl( \frac{\epsilon}{2} \bigr) \log \frac{H^2}{4\pi}+\mathcal{O}(\epsilon^2) \Bigr)\Bigl(-\frac{2}{\epsilon}-1+\gamma+\mathcal{O}(\epsilon) \Bigr) \\
&\ \ \ \  \frac{m^2}{H^2} \biggl[ 1-\Bigl(\frac{\epsilon}{2}\Bigr) \Bigl( \psi(1+s)+\psi(1-s)+\frac{1}{1-s} \Bigr)+\mathcal{O}(\epsilon^2) \biggr], \\
&=\frac{m^2}{16 \pi^2} \Bigl( -\frac{2}{\epsilon}-1+\gamma+\mathcal{O}(\epsilon) \Bigr) \\
 &\ \ \ \  \biggl[1-\Bigl(\frac{\epsilon}{2}\Bigr) \Bigl( \psi(1+s)+\psi(1-s)+\frac{1}{1-s}+\log\frac{H^2}{4 \pi} \Bigr)+\mathcal{O}(\epsilon^2) \biggr], \\
&=\frac{m^2}{16 \pi^2} \biggl[-\frac{2}{\epsilon}+\Bigl( \psi(1+s)+\psi(1-s)+\frac{1}{1-s}-1+\gamma+\log\frac{H^2}{4 \pi} \Bigr)+\mathcal{O}(\epsilon) \biggr]. 
\end{split}
\end{equation}
Again, if we expand $s$ around the massless case assuming $m^2/H^2 \ll 1$, we obtain
\begin{equation}
\begin{split}
G(x,x)=&-\frac{m^2}{16 \pi^2} \frac{2}{\epsilon}+\frac{m^2}{16 \pi^2} \bigl( \gamma+\log\frac{H^2}{4 \pi} \bigr) \\
&+\frac{m^2}{16 \pi^2} \Bigl( -2 \gamma+1-4 \frac{m^2}{H^2} \Bigr)+\mathcal{O} \bigl(\epsilon, (\tfrac{m^2}{H^2})^2 \bigr).
\end{split}
\end{equation}
Note that in the conformally coupled case, the divergence structure of the tadpole correction is the same as that in flat space.

\subsection{General case}
For generality we denote the both coincident propagators as follows 
\begin{equation}
\begin{split}
G(x,x)=(m^2+\kappa H^2) T_d+T_F,
\end{split}
\end{equation}
where $T_d=-2/16\pi^2 \epsilon$. In the minimally coupled case for the small mass expansion, the expressions for $\kappa H^2$ and $T_F$ are given by
\begin{equation}
\kappa H^2=-2 H^2,
\end{equation}
\begin{equation}
T_F= \frac{1}{16 \pi^2} (m^2-2 H^2) \bigl( \gamma+\log\frac{H^2}{4 \pi} \bigr)
             +\frac{H^2}{16 \pi^2} \biggl[ \frac{6 H^2}{m^2}+4\gamma-\frac{23}{3}-\bigl( 2\gamma+\frac{2}{27} \bigr) \frac{m^2}{H^2} \biggr]+\mathcal{O} \bigl(\epsilon, (\tfrac{m^2}{H^2})^2 \bigr).
\end{equation}
In the conformally coupled case for the small mass expansion, these expressions are given by
\begin{equation}
\kappa H^2=0,
\end{equation}
\begin{equation}
T_F=\frac{m^2}{16 \pi^2} \bigl( \gamma+\log\frac{H^2}{4 \pi} \bigr)
+\frac{m^2}{16 \pi^2} \Bigl( -2 \gamma+1-4 \frac{m^2}{H^2} \Bigr)+\mathcal{O} \bigl(\epsilon, (\tfrac{m^2}{H^2})^2 \bigr).
\end{equation}

\bibliography{basename of .bib file}

\end{document}